\documentstyle[12pt,aasms4]{article}
\singlespace

\slugcomment{Submitted to the Astronomical Journal}

\lefthead{}
\righthead{}

\begin{document}

\title{Pulsar wind nebulae around the southern pulsars PSR B1643$-$43 and 
PSR B1706$-$44}

\author{E. B. Giacani\altaffilmark{1}}
\affil{Instituto de Astronom\'\i a y F\'\i sica del Espacio (CONICET,
  UBA),}
\centerline{C.C.67, 1428 Buenos Aires, Argentina.}
\centerline{e-mail: egiacani@iafe.uba.ar}

\author{D. A. Frail}
\affil{ National Radio Astronomy Observatory, P.O.Box 0,}
\affil{ Socorro, New Mexico 87801, USA.}
\centerline{e-mail: dfrail@nrao.edu}

\author{W. M. Goss}
\affil{National Radio Astronomy  Observatory, P.O. Box 0,}
\affil {Socorro, New Mexico 87801, USA.}
\centerline {e-mail: mgoss@nrao.edu}

\and

\author{M. Vieytes}
\affil{Instituto de Astronom\'\i a y  F\'\i sica del Espacio, CONICET,UBA}
\centerline{C.C.67, 1428 Buenos Aires, Argentina}
\centerline{e-mail:mary@sion.com}

\noindent {\bf Address to send proofs:} 
E.Giacani, Instituto de Astronom\'\i a y F\'\i sica del Espacio,
C.C.67, Suc 28, 1428 Buenos Aires, Argentina

\altaffiltext{1}{Member of the Carrera del Investigador
Cient\'\i fico of CONICET, Argentina}

\begin{abstract}
  We present high resolution VLA images taken at the wavelengths of
  $\lambda 20$ cm, $\lambda 6$ cm, and $\lambda 3.6$ cm in the
  vicinity of the pulsars PSR B1706$-$44 (PSR J1709$-$4428) and PSR
  B1643$-$43 (PSR J1646-4346).  Both of these pulsars are young
  ($<$30,000 yrs) and have large spin-down luminosities (\.E$>10^{35}$
  ergs$^{-1}$) and hence are good candidates to search for extended
  synchrotron nebula excited by the relativistic pulsar wind. For PSR
  B1643$-$43 we found evidences of a 4$^\prime$ comet-shaped nebula,
  suggestive of a synchrotron ``wake'' left by a fast moving pulsar.
  PSR B1706$-$44 appears surrounded by a spherical nebula  approximately
  3$^\prime$ in diameter. Based on their morphology, the detection of
  significant linear polarization ($>$20\%), and their flat radio
  spectra ($\alpha=0.25-0.3$, where S$_\nu\propto\nu^{-\alpha}$) we
  argue that these are wind nebulae powered by the rotational energy
  loss of the respective pulsars.
\end{abstract}

\clearpage

\section{Introduction}\label{sec:intro}

Pulsars transfer the bulk of their rotational angular momentum in a
wind of relativistic particles and Poynting flux (Michel 1969, Rees \&
Gunn 1974, Kennel et al. 1983).  However, because the particles emerge
from the magnetosphere with zero pitch angle, the winds cannot be
directly observed.  Therefore in order to study the winds from pulsars
it is necessary to study the synchrotron emission from the pulsar wind
nebula (PWN) produced when the wind is thermalized as it comes into
pressure equilibrium with the surroundings. The Crab nebula is the
archetype of the PWN; this source has been studied in detail at all
wavelengths (e.g. X-ray: Brinkmann, Aschenbach \& Langmeier 1985,
radio: Bietenholz \& Kronberg 1992, optical: Hester et al.  1995). The
fact that the bolometric luminosity of the Crab nebula requires an
on-going source with a power comparable to the rotational energy loss
(\.E) of the pulsar, shows that the PWN owes its existence to the
existence of a young, energetic pulsar.

At radio wavelengths, there are at least two morphological types of
PWN, depending on the source of confinement for the wind (Frail \&
Scharringhausen 1997, Chevalier 1998, Gaensler et al. 2000): the so
called ``filled-center'' and the ``bow-shock''. The
filled-center PWN or ``plerions'' (e.g. Crab) are inside young
supernova remnants and are confined by the hot gas driving the
expansion. The bow-shock PWN are found both inside and outside
supernova remnants (e.g. PSR B1757$-$24 and G5.4$-$1.2; Frail \&
Kulkarni 1991) and are confined by the high space velocities of the
pulsar. The non-thermal radio emission from PWN can be distinguished
from shock-accelerated emission in supernova remnants by (1) a flat
spectrum, $\alpha=0.1$ to $0.3$, where S$_\nu\propto\nu^{-\alpha}$, and
(2) a high degree of linear polarization ($>>5\%$) (Chevalier 1998).

The existing sample of radio PWN is small. There are no more than
seven confirmed PWN at radio wavelengths (the number is comparable for
X-ray PWN) which contain a known pulsar. Frail, Goss \& Whiteoak
(1994) (hereafter FGW94) made radio images around three young pulsars,
and identified two promising PWN candidates based on their morphology
alone. In this paper we present multi-frequency polarimetric
observations made with the National Radio Astronomy Observatory
(NRAO\footnotemark\footnotetext{The NRAO is a facility of the National
  Science Foundation operated under cooperative agreement by
  Associated Universities, Inc.}) Very Large Array (VLA) toward PSR
B1643$-$43 and PSR B1706$-$44 to better ascertain the properties of
the emission in the vicinity of these pulsars.

\section {Observations} 

The extended emission in the vicinity of PSR B1643$-$43 and PSR
B1706$-$44 was imaged at 1425, 4860 and 8460 MHz in several observing
runs during 1997, using different configurations of the VLA (see Table
1).  The data were obtained in the Stokes parameters I, Q, U, and V.
The {\it uv} data from each array were combined to form a single
dataset for each pulsar in order to image a full range of spatial
frequencies. An additional 1.4 GHz dataset, taken in 1993 with the VLA
in its CnB and DnC hybrid configurations, was extracted from the
archive. A description of these data can be found in FGW94.  All data
reduction and calibration were done following standard practice in use
at the VLA. The images were corrected for primary beam. Table 1
summarizes the observational parameters.

\section {Results }

\subsection{PSR B1643$-$43}

PSR B1643$-$43 (PSR J1646-4346) is a 232 ms pulsar, with a
characteristic age of 32.6$\times{10}^3$ yr, a spin-down luminosity 3.6
$\times 10 ^{35}$ erg s$^{-1}$ (Johnston et al. 1995) and a dispersion
measure-based distance of 6.9 kpc (Taylor \& Cordes 1993).  A search
for pulsed $\gamma$-ray emission was made with the EGRET and COMPTEL
instruments on board of the Compton Gamma-Ray Observatory with null
results (Thompson et al. 1994, Carrami\~nana et al.~1995).

Based on observations carried out with the VLA in the radio continuum at  $\lambda 20$ cm 
in an extended  region around PSR
B1643$-$43, FGW94 have shown   that the pulsar is located within the shell of the
SNR G341.2+0.9, about 8$^\prime$ west of the center of the remnant
(see Fig.1{ \it Left}).  Morphological evidence and a coincidence in distance led
FGW94 to propose a physical association between the pulsar and the
remnant.  FWG94 have also reported the detection of a 4$^\prime$
nebulosity with a cometary morphology just east of PSR B1643$-$43
which is joined to the pulsar by a bridge of emission.  Based on
morphological evidence the authors suggested that this feature is the
synchrotron nebula left behind by the fast moving pulsar. The characteristic age of PSR B1643-43
and its positional offset from the center of G341.2+0.9 together imply a transverse
velocity of 475 km s$^ {-1}$, which in the absence of proper motion measurements predicts
that the pulsar would have to be moving about 15 mas yr$^{-1}$ (FGW94).

In the same survey, FWG94 noted that the position of PSR B1643$-$43
most likely coincided with a 1.5 mJy point source detected in their
$\lambda 20$ cm VLA radio image. Their interferometric position
differed significantly (40\arcsec) from the timing position of the
pulsar (Johnston et al.~1995). The later is probably an error due
to the timing noise and ``glitches'' that characterize the spindown of
young pulsars.  The present observations, performed with better
sensitivity and spatial resolution than in FGW94, allow us to derive an
improved interferometric position for the pulsar.  Our best position
determination is from the BnA array $\lambda 20$ cm data with a
beamwidth of 7$^{\prime\prime}.3 \times 3^{\prime\prime}.7$, a
considerable improvement over the 25$^{\prime\prime}$ beam in FGW94. A
fit to the peak gives a position (epoch B1950) of R.A.= $16^{h}43^{m}
16^{s}.56 \pm 0^{s}.04$, decl.=$-43^{\circ} 40^\prime
31^{\prime\prime}.8 \pm 0^{\prime\prime}.8$, or (epoch J2000) R.A.=$
16^{h}46^{m} 50^{s}.86 \pm 0^{s}.04$, decl.=$-43^{\circ} 45^\prime
53^{\prime\prime} .7 \pm 0^{\prime\prime}.8$. We have also determined
the position of the pulsar using each dataset separately and at all
frequencies and find good agreement. The derived position is about
5$^{\prime\prime}$ northwest of the position given by FWG94. We
suspect that extended nebular emission underlying the pulsar is
shifting the centroid and therefore the present position with its much
smaller beam is superior to the that of FGW94.

Figure 1 ({\it Right}) shows a grayscale and countour images of a region surrounding
PSR 1643$-$43 at 1.4 and 4.8 GHz. These images were convolved to a
25\arcsec\ circular beam. The cross indicates the new position of the
pulsar. Because of the incomplete sampling of the visibility plane,
the VLA image at 8.4 GHz lacks short spacings and is not included
here.

From Figure 1 ({\it Right}) a synchrotron nebula of about 3$^\prime$ in size is
clearly visible.  The nebula, with the pulsar located at its western
border, appears as a feature pointing back from the pulsar to the
center of G341.2+0.9 in the direction opposite the pulsar's implied
proper motion.  Such a morphology is compatible with the
interpretation of this structure as a PWN. Similar radio morphologies
have been detected before surrounding PSR B1757-24 in the SNR
G5.4$-$1.2 (Frail \& Kulkarni 1991), and PSR B1853+01 in W44 (Frail et
al.~1996).

We find additional evidence to support our contention that this
feature is the PWN associated with PSR B1643$-$43. i) We detect
significant linearly polarized intensity at 4.8 and 8.4 GHz, with a
mean fractional polarization of about 30\%. ii.) We also estimated the
total flux density for the nebular emission (and for the pulsar).
These values are given in Table 2. The errors quoted include the rms
noise of each image and the uncertainty in the choice of the
integration boundaries. From a least squares fit we derive a radio
spectral index $\alpha$ = 0.24 between 330 MHz and 4.8 GHz, where the
data at 330 MHz were taken from FWG94 and data at 843 MHz were taken
from the MOST survey by Green et al. (1999). A high degree of linear
polarization and a flat radio spectral index are two unmistakable
properties of PWN in the radio band, and thus we propose that this is
the synchrotron nebula excited by the pulsar wind.

For a distance of 6.9 kpc, the corresponding radio luminosity L$_{r}$
of the PWN between 10$^7$ and 10$^{11}$ Hz is 8.3 $\times 10^{31}$ erg
s$^{-1}$. This value corresponds to an efficiency $\epsilon \equiv$
L$_{r}$/\.E = 1.6 $\times 10^{-4}$, in very good agreement with the
$\epsilon \sim 10^{-4}$ derived by Gaensler et al. (2000) for young
energetic pulsars.

\subsection {PSR B1706$-$44 }

PSR B1706$-$44 (PSR J1709$-$4428) is a young pulsar (spin-down age
$\sim 17000$ yrs), with a period of 102 ms and a large spin-down
luminosity of $3.4 \times 10^{36}$ ergs$^{-1}$.  It is also one of
a very small number (6) of radio pulsars to have been detected as a
pulsed gamma-ray source (Thompson et al. 1992). There are several
lines of evidence that suggest the existence of a filled-center nebula
surrounding the pulsar. FGW94 have noted in a low resolution
(24\arcsec) radio image at $\lambda 20$ cm that PSR B1706$-$44 appears
embedded in a ``halo" about 4$^\prime$ in size . The authors suggest
that the emission could be due to a PWN around the pulsar.  In the
soft X-ray band, between 0.1-2.4 keV, unpulsed radiation was detected,
with a 2 $\sigma$ upper limit on the pulsed fraction of the 18 \%.
This unpulsed emission is thought to originate from a compact
synchrotron nebula of about 1$^\prime$ in size around the pulsar
(Becker et al.~1995, Finley et al. 1998). More recent X-ray
observations made with {\it ROSAT}, {\it ASCA} (Finley et al.~1998)
and {\it RXTE} (Ray et al. 1998) confirm the lack of pulsations.
Observations of PSR B1706$-$44 in the very high energy $\gamma$ rays
by the CANGAROO imaging Cerenkov telescope have revealed unpulsed TeV
radiation at a 10 $\sigma$ confidence level (Kifune et al. 1995). It
was suggested that the TeV emission could be due to inverse Compton
radiation from a PWN (Harding 1996, Aharonian et al.  1997).
Chakrabarty \& Kaspi (1998) have reported negative results of a search
for optical pulsations from PSR B1706$-$44. Sefako et al.  (2000) have
carried out V band CCD observations in the direction of the pulsar in
order to look for the optical counterpart of the 1$^\prime$ compact
X-ray nebula, but their search did not reveal any nebular structure
around the pulsar.

A possible association between the supernova remnant (SNR) G343.1-2.3
and PSR B1706$-$44 was proposed by McAdam et al.~(1993). Such an
association, however, was questioned by FGW94 and Nicastro, Johnston
\& Koribalski (1996) based on distance inconsistences, a lack of
morphological signatures of interaction between the pulsar and the
SNR, and scintillation measurements indicating a transverse velocity
for the pulsar at least 20 times smaller than required if the pulsar
originated in the geometrical center of G343.1-2.3, 17000 years ago.

A dispersion based distance measure of Taylor \& Cordes (1993)
places PSR B1706$-$44 at 1.8 kpc; while HI absorption shows that its
distance lies in the range 2.4-3.2 kpc (Koribalski et al. 1995). In
what follows we will adopt a distance to the pulsar of 2 kpc.

As in the previous case for PSR B1643$-$43, we have determined the
position of the pulsar using all data sets separately. Here again, the
most accurate fit was obtained from the high angular resolution of the
BnA array $\lambda 20$ cm array data (beamwidth 9$^{\prime\prime}.1
\times 4^{\prime\prime}.8$). The derived position is (epoch B1950)
R.A.= $17^{h} 6^{m} 5^{s}.09 \pm 0^{s}.02$, decl.=$-44^{\circ}
25^\prime 20^{\prime\prime}.6 \pm 0^{\prime\prime}.5$, or (epoch
J2000) R.A.=$ 17^{h} 9^{m} 42^{s}.75 \pm 0^{s}.02$, decl.=$-44^{\circ}
29^\prime 6^{\prime\prime} .6 \pm 0^{\prime\prime}.5$. This position
is in good agreement with the interferometric measurements of FGW94
and the new timing position by Wang et al.~(2000).

Figure 2 shows  greyscale and contour images of the region
surrounding the pulsar at 1425, 4860 and 8640 MHz. These images were
convolved with a circular beam of 25$^{\prime\prime}$. The cross indicates
the position of PSR B1706$-$44.  The pulsar appears surrounded by a
synchrotron nebula, about 3'.5$\times$ 2'.5 in size, with the
brightest part towards the east.

The total flux density of the nebular emission and PSR B1706$-$44 is
summarized in Table 2. Again, the quoted errors take into account
uncertainties in the definition of the outer boundaries. From a least
squares fit we derive a radio spectral index $\alpha$ = 0.3 between
330 MHz and 8.4 GHz, where the data at 330 MHz were taken from FWG94.
Significant linearly polarized intensity was detected at 4.8 and 8.4
GHz, with a mean fractional polarization of about 20 \%. This is
convincing evidence that we have detected another PWN.
 
The radio luminosity L$_{r}$ of the PWN between 10$^7$ and 10$^{11}$
Hz is: 7.6$\times 10^{30}$ ergs$^{-1}$, corresponding to an efficiency
$\epsilon \equiv$ L$_{r}$/\.E $\approx$ 2$\times{10}^{-6}$. These
values are significantly lower than for any other radio PWN (Frail \&
Scharringhausen 1997). The equipartition magnetic field in the nebula
can be estimated by the usual means (Pacholczyk 1970), assuming that
the energy density of the magnetic field is half of the total
synchrotron pressure. For an electron/positron plasma with unity
volume filling factor we find B$_{eq}=20$ $\mu$G and a minimum energy
of 3$\times{10}^{45}$ erg.

In order to match the observed X-ray and $\gamma$-ray fluxes in PSR
B1706$-$44, Harding (1996) proposed a scenario where the TeV emission
is produced within the synchrotron nebula via the IC mechanism and the
target photon field is the 2.7 K microwave background radiation (MBR).
The author abtains a good fit if the magnetic field strength inside the
nebula is lower than 5 $\mu$G, value lower than our estimation.

On the other hand, Aharonian et al. (1997) proposed that the
production of TeV $\gamma$-rays take place in a region of about $0.1
^{\circ}$, outside of the compact X-ray nebula.  Inside the compact
nebula, where synchrotron is the dominant process, the magnetic field
takes values between 20 to 60 $\mu$G (depending on the model
parameters); out of this region, where the MBR photons are accelerated
to TeV energy by the IC mechanism, the value of the magnetic field is
about 3 $\mu$G.

Our estimate of the magnetic field in the radio nebula is in good
agreement with that of the Aharonian et al's model for the compact
X-ray nebula. However, this new radio observations indicate that the
synchrotron nebula extends up to 3'.5. The $\gamma$ photons would
therefore need to be produced even farther out.

Finley et al. (1998) have also explained the unpulsed TeV emission in
PSR B1706$-$44 as originated by the IC mechanism but in their
model, the target photon
field is the infrared background radiation. Their results are
consistent with a continuous unbroken power-law spectrum extending
from radio to X-ray domain. However, the curent radio data argue against this model.
Our observations allow to  better define the radio
spectrum  producing an $\alpha \simeq 0.3$, which when combined with
that obtained from the X-ray observations implies the existence of at
least one break in the synchrotron
spectrum between the radio and X-ray bands.
 
\section {Conclusions}

We have made radio observations toward PSR B1643$-$43 and PSR
B1706$-$44 and found good evidences that these pulsars are surrounded
by extended emission, powered by their winds. At the time of the
survey of Frail \& Scharringhausen (1997) only six PWN were known.
This present work and Gaensler et al. (1998) has extended this sample
by 50\%. These new radio PWN have the same properties of the rest of
the sample (i.e. morphology, spectra, polarization) as a whole with
one exception. Most radio PWN radiate of order 10$^{-4}$ of their
spindown luminosity but there are a large number of non-detections
(Gaensler et al. 2000) suggesting that this ratio is not constant and
may be in fact much lower in specific cases. PSR B1706$-$44 and PSR
B0906$-$49 both have a PWN with $\epsilon$ ($\simeq2\times{10}^{-6}$)
much lower than the rest of the sample but comparable to those
inferred from the upper limits of other young pulsars. In the 
case of PSR B0906$-$49, the spectrum of the radio nebula is steeper
than other radio PWN and the pulsar is the older than   any other 
pulsar known to power a radio PWN.  Clearly we need
further broad-band studies of PWN produced under a variety of conditions 
to better understand  how the radio emission from a PWN is produced and
how it depends on the properties of its pulsar.

\vskip 3cm

{\bf Acknowledgments}

This research was funded by a Cooperative Science Program
between the National Science Foundation and CONICET (Argentina) and
through the CONICET grant 4203/96 and ANPCYT grant 0300000-0235.

\newpage
\centerline {\bf Figure Captions}

\figcaption{{\it Left:} VLA continuum image of the SNR G341.2+0.9
at $\lambda$20cm from Frail, Goss \& Whiteoak (1994). The box indicates the
extended radio emission surrounding PSR B1643-43. {\it Right and Top:}
 Greyscale and contour image of the region
  surrounding PSR B1643$-$43 at 1425 MHz. The plotted contours are
  4.2, 4.5, 4.8, 5.1, 5.4, 5.7 and 6.0 mJy/beam. The beam size is
  25\arcsec. The greyscale ranges from 4 to 5.8 mJy/beam. The cross
  indicates the position of the pulsar as derived in this paper.  {\it
Right and   Bottom:} The same region at 4860 MHz. Contours are at levels of
  0.16, 0.3, 0.4, 0.5, 0.7, 0.9 and 1.1 mJy/beam. The greyscale ranges
  from 0 to 1.3 mJy/beam. The cross indicates the position of the
  pulsar. }

\figcaption{{\it Left and Top:} Greyscale and contour image of the
  region surrounding PSR B1706$-$44 at 1425 MHz. The plotted contours
  are 2.1, 2.25, 2.4, 2.55, 2.7, 3.0, 6.0 and 9.0 mJy/beam. The beam
  size is 25\arcsec. The greyscale goes linearly from 2.1 to 2.9
  mJy/beam. The cross indicates the position of the pulsar as derived
  in this paper.  {\it Right and Top:} The same region at 4860 MHz.
  The contours are 0.1, 0.3, 0.5, 0.7, 0.9, 1.1, 1.3 and 1.5 mJy/beam.
  The beam size is 25\arcsec. The greyscale ranges from 0 to 1
  mJy/beam. The cross indicates the position of PSR B1706$-$44.  {\it
    Bottom:} The same region around PSR B1706$-$44 at 8460 MHz.  The
  contours level are 0.1, 0.3, 0.5, 0.65, 0.9 and 1.1 mJy/beam. The
  greyscale ranges from 0 to 0.7 mJy/beam. The cross indicates the
  position of the pulsar.}
  
\newpage
\begin{deluxetable}{lccc}
\tablecaption{Observational Parameters\label{tbl1}}
\startdata
&\nl
\tableline
\tableline
Frequency [MHz]:&1425& 4860& 8460\nl
VLA configuration:&BnA + CnB + DnC & CnB + DnC & CnB + DnC\nl
Observing dates:&1997 Feb.7,8&1997 June 15& 1997 June 16,17\nl
&1993 June 2, 1997 June 13& 1997 Oct.14& 1997 Oct. 6,10\nl
&1993 Oct. 14, 1997 Oct. 17\nl
Total observing time:&7+17.5+14 h& 17.5+14 h& 17.5+14h\nl
Calibrators:&3C286,  1622-297&3C286, 1622-297&3C286, 1622-297\nl
Synthesized beam [arcsec]:\nl
PSR B1643$-$43:& 27$\times 10$ & 16$\times 6$ & 19.4$\times 14 $\nl
PSR B1706$-$44:&$24\times 9$&16$\times 6$& 14 $\times 8$\nl
rms [mJy]:& 0.04  &  0.04 & 0.04 \nl
\enddata

\end{deluxetable}

\newpage
\begin{deluxetable} {rrll}
  \tablecaption{Observed parameters of the pulsar wind nebulae}
\tablehead{
\colhead{Pulsar}  &
\colhead{$\nu$ (GHz)} &
\colhead{S$_{PSR}$ (mJy)} &
\colhead {S$_{PWN}$ (mJy)} \\ 
}
\startdata
B1643$-$43 & 1.42 & 1.0 $\pm 0.2 $ & 31 $\pm 13 $  \\
& 4.86  &  0.10 $\pm 0.03$ &  23 $\pm 4$  \\
B1706$-$44 & 1.42 & 11.0 $\pm 0.2 $ & 28$\pm 11$ \\
& 4.86 & 2.0$\pm 0.7$ & 28$\pm 6 $ \\
& 8.46 & 0.8 $\pm 0.1$ & 11$\pm 3$
\enddata
\tablecomments{The columns are (left to right), (1) the pulsar's name,
 (2) frequency, (3) flux density of the pulsar,  and (4) integrated
flux density of the PWN.}
\end{deluxetable}

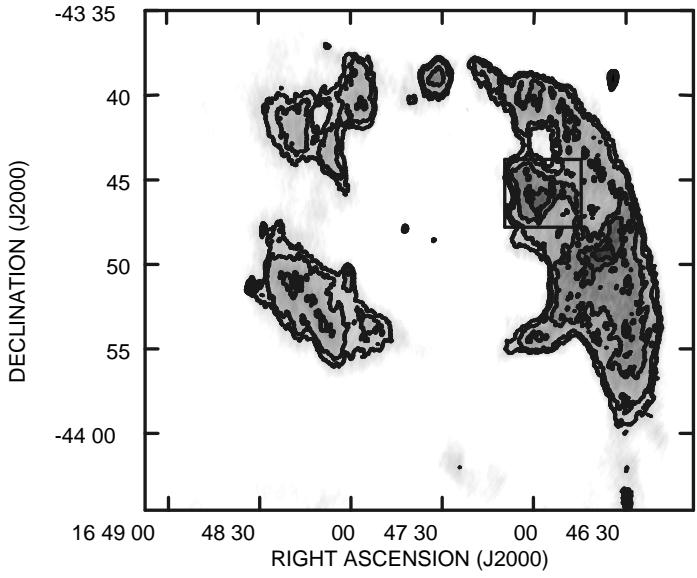
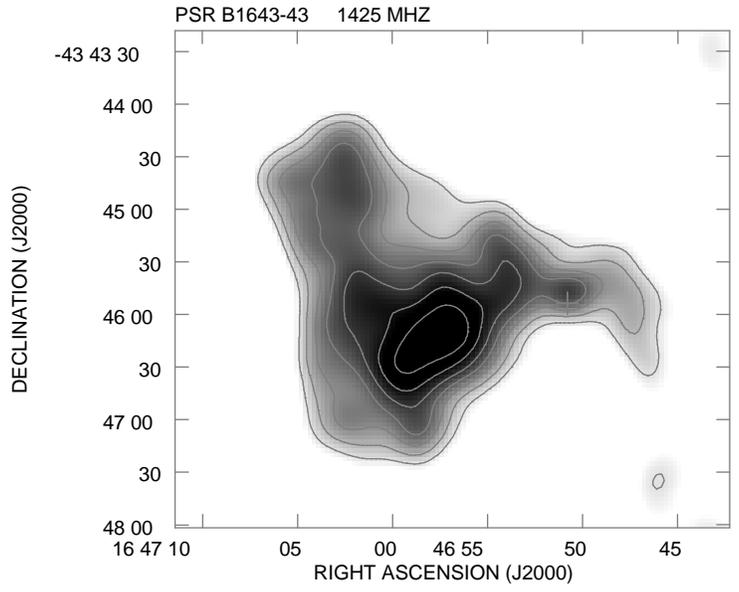
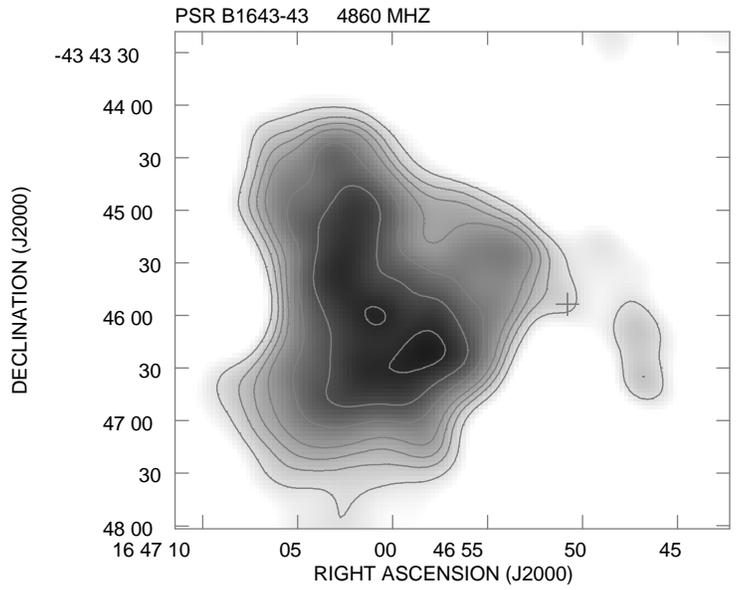
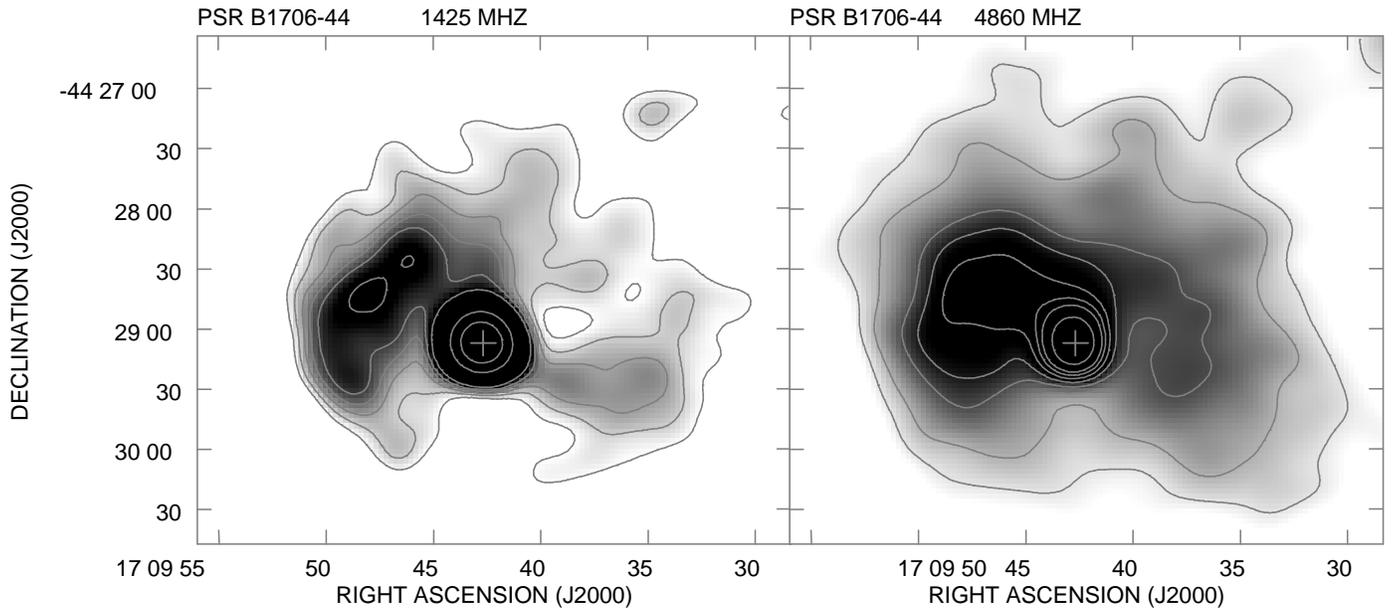
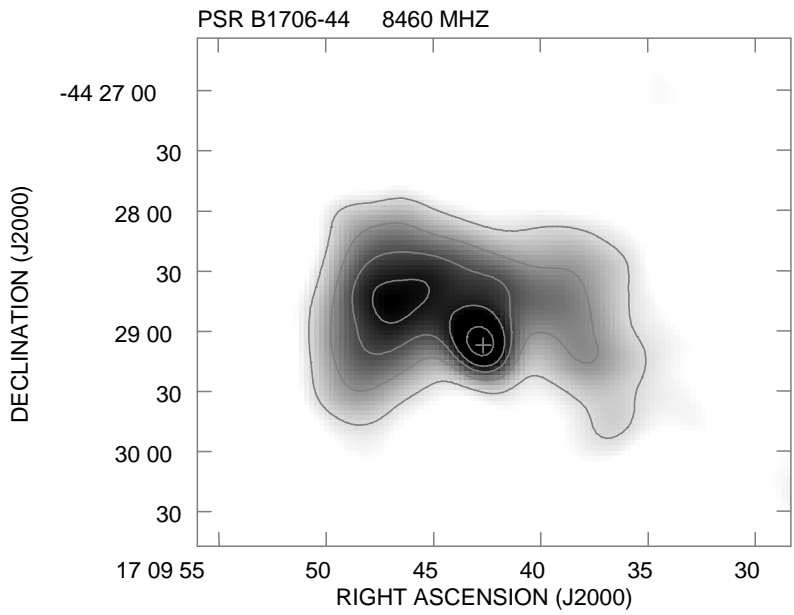
\end{document}